\documentclass[prl,twocolumn,showpacs,superscriptaddress,preprintnumbers,amsmath,amssymb]{revtex4-1}
\usepackage{graphicx,color}
\usepackage{dcolumn}
\usepackage{bm}
\usepackage{epstopdf}
\usepackage{tabularx}

\begin{document}

\preprint{APS/123-QED}

\title{Angular momentum transfer via relativistic spin-lattice coupling
  from first principles. }

\author{Sergiy Mankovsky}
\affiliation{%
Department of Chemistry/Phys.\ Chemistry, LMU Munich,
Butenandtstrasse 11, D-81377 Munich, Germany 
}%
\author{Svitlana Polesya}
\affiliation{%
Department of Chemistry/Phys.\ Chemistry, LMU Munich,
Butenandtstrasse 11, D-81377 Munich, Germany 
}%
\author{Hannah Lange}
\affiliation{%
Department of Chemistry/Phys.\ Chemistry, LMU Munich,
Butenandtstrasse 11, D-81377 Munich, Germany 
}%
\author{Markus Wei{\ss}enhofer}
\affiliation{%
Department of Physics, University of Konstanz, DE-78457 Konstanz, Germany 
}
\author{Ulrich Nowak}
\affiliation{%
Department of Physics, University of Konstanz, DE-78457 Konstanz, Germany 
}%
\author{Hubert Ebert}
\affiliation{%
Department of Chemistry/Phys.\ Chemistry, LMU Munich,
Butenandtstrasse 11, D-81377 Munich, Germany 
}%

\newcommand{\BLUE}{\bf \color{blue} }
\newcommand{\RED}{\bf \color{black} }
\newcommand{\DONE}[1]{ \marginpar{\bf\large DONE} {\bf \em #1}  \marginpar{\bf\large DONE}}

\newcommand{\DISC}{ \marginpar{\bf\em\large DIS\-CUS\-SION} }  

\date{\today}

\begin{abstract}
The transfer and control of angular momentum is a key aspect for spintronic applications. Only recently, it was shown that it is possible to transfer angular momentum from the spin system to the lattice on ultrashort time scales. In an attempt to contribute to the understanding of angular momentum transfer between spin and lattice degrees of freedom we present a scheme to calculate fully-relativistic spin-lattice coupling parameters from first-principles.
By treating  changes in the spin configuration and atomic positions at the same level, 
closed expressions for the atomic spin-lattice coupling parameters can be derived in a coherent manner up to any order.
Analyzing the properties of these parameters, in particular their dependence on spin-orbit coupling, we find that even in bcc Fe the leading term for the angular momentum exchange between the spin system and the lattice is a Dzyaloshiskii-Moriya-type interaction, which is due to the symmetry breaking distortion of the lattice. 
\end{abstract}

\pacs{71.15.-m,71.55.Ak, 75.30.Ds}
\maketitle

Spintronics is an emerging field aiming for the development of future nanoelectronic devices. A key aspect is the transport and control of angular momentum \cite{Hir20}. While the focus has long been on spin-polarized electrons to
carry the angular momentum, newer lines of research include the magnonic
spin as angular momentum carrier, opening perspectives for insulator
spintronics. Understanding the flow of angular momentum is also vital
for the progress of ultrafast magnetization switching. In certain
ferrimagnets, a single laser pulse can switch the magnetization
orientation on a sub-picosecond time scale \cite{Rad11}, due to the
exchange of spin angular momentum between the two magnetic sublattices
\cite{Men12,WHC+13}. However, recent work on ultrafast demagnetization in
ferromagnets has demonstrated that angular momentum can also be
transferred from the spin system to the lattice on similar time scales
\cite{Tau22}. In the lattice, the spin angular momentum is absorbed in
terms of phonons carrying the angular momentum till --- on larger times
scales --- the macroscopic Einstein-deHaas effect sets in \cite{DAS+19}.   
These findings add another piece to the mysteries of spintronics and
ultrafast phenomena, namely the understanding of the microscopic
mechanisms that transfer angular momentum between the spin system and
the lattice. 

The calculation of spin-lattice coupling terms, including the exchange
of angular momentum between spins and lattice degrees of freedom, is
only at its beginning \cite{GC15,MKL19,SVSB19,RSBD20}. The development
of new tools for the quantitative calculation of spin lattice dynamics
--- so-called molecular-spin dynamics simulations
\cite{MWD08,PEN+16,AN19,HTM+19,SER+21} --- is delayed by the fact that a
systematic derivation of proper spin-lattice parameters is still
missing. For a pure spin model the calculation of exchange coupling
parameters $ J_{ij} $ of the Heisenberg Hamiltonian on a
first-principles level by means of the so-called Lichtenstein formula
\cite{LKAG87} is now a well established approach to supply the necessary
input for 
Monte Carlo \cite{PMS+10} as well as spin-dynamics simulations
\cite{Mry05,SHNE08}. Corresponding extensions of this computational scheme are
now available to account for the full tensorial form of the interaction
parameters \cite{USPW03,EM09a} as well as their extension to a
multi-site  formulation  \cite{MPE20}. 

Including the lattice degrees of freedom is much more challenging.  
A practical scheme to calculate microscopic spin-lattice coupling (SLC)
parameters quantitatively and on the basis of  electronic structure
calculations has been suggested so far only by  Hellsvik et al.\
\cite{HTM+19} by applying the  
   Lichtenstein formula as well as its relativistic generalization \cite{SBK+22} for a system with one atom moved gradually from
   its equilibrium position. 

In this letter we present and exploit an improved, fully-relativistic
scheme that treats changes to the spin configuration and atomic
positions on the same level. This allows to derive closed expressions
for the atomic SLC parameters in a coherent way up to any order.
First
numerical results are presented for the three-site terms of bcc
Fe. Surprisingly, even in a bcc crystal the leading term for the
exchange of spin angular momentum with the lattice is a
Dzyaloshiskii-Moriya-type interaction emerging due to the symmetry
breaking distortion of the lattice.

To describe  the coupling of spin and spatial degrees of freedom we adopt an atomistic
approach and start with the expansion of a phenomenological spin-lattice Hamiltonian 
\begin{eqnarray}
{\cal H}_{sl} &=&  - \sum_{i,j,\alpha,\beta} J_{ij}^{\alpha \beta}
                   e_i^{\alpha}e_j^{\beta} 
  -  \sum_{i,j,\alpha,\beta} \sum_{k,\mu} J_{ij,k}^{\alpha
     \beta,\mu} 
                   e_i^{\alpha}e_j^{\beta}  
                   u^{\mu}_k
\nonumber \\ 
  &&  
\qquad \qquad \quad  
  -  \sum_{i,j}\sum_{k,l}  J_{ij,kl}^{\alpha
     \beta,\mu\nu}  
                   e_i^{\alpha}e_j^{\beta}  
                    u^{\mu}_k u^{\nu}_l
                   \;,
\label{eq:Hamilt_extended_magneto-elastic}
\end{eqnarray}
%
that can be seen as a lattice extension of a Heisenberg spin Hamiltonian.
Accordingly, the spin and lattice degrees of freedom
are represented by the orientation vectors $\hat{e}_i$ of the
magnetic moments $\vec m_i$ and displacement  vectors $\vec u_i$ for each 
atomic site $i$. In Eq.\ (\ref{eq:Hamilt_extended_magneto-elastic})
we omit pure lattice terms that involve the force constants \cite{HTM+19}
as we focus here on the magnetic part of the Hamiltonian.
The pure spin part (term 1) has been restricted to its two-site
contributions with the corresponding coupling parameters denoted
two-site SSC below. An extension of this term to higher order can be
done in a straight forward manner \cite{MPE20}. 
Also, the spin-lattice coupling has been restricted to three and four-site
terms (terms 2 and 3).
As  relativistic effects are taken into account, the exchange interactions are described
in  tensorial form, $J_{ij}^{\alpha \beta}$, instead of the simpler
isotropic form $J_{ij}^{}$ of the standard Heisenberg spin Hamiltonian.
The Hamiltonian in Eq.\ (\ref{eq:Hamilt_extended_magneto-elastic})
is similar in form to the one discussed by Hellsvik et al. \cite{HTM+19}, 
potentially providing a suitable basis for advanced molecular-spin dynamics simulations.

In previous works expressions for  the exchange coupling parameters
$J_{ij}$  \cite{LKAG87} or $J_{ij}^{\alpha\beta}$ \cite{USPW03,EM09a}, respectively,   
have been derived by mapping the  free energy landscape ${\cal F}(\{\hat{e}_i\})$
obtained from first-principles electronic structure calculations
on the Heisenberg spin Hamiltonian.
Here, we follow the same strategy by mapping the
free energy landscape ${\cal F}(\{\hat{e}_i\},\{\vec u_i\})$ accounting for its dependence on the spin configuration $\{\hat{e}_i\}$ as well as  atomic displacements $\{\vec u_i\}$ on the same footing.
Making use of the magnetic force theorem the change in free energy $ \Delta {\cal F} $
induced by changes of the spin configuration  $\{\hat{e}_i\}$ with respect to a suitable reference system
and simultaneous finite  atomic displacements $\{\vec u_i\}$ can be written in terms of corresponding changes
to the single-particle energies,
%
\begin{eqnarray}
  \Delta {\cal F} =  \int^{E_F} \! \! \! dE (E - E_F) \Delta n(E) = - \! \! \int^{E_F}  \! \! \! dE \Delta N(E) \, , \;
\label{eq:Free_energy}
\end{eqnarray}
where $E_F$ is the Fermi energy and $ \Delta n(E)$
and  $\Delta N(E)$ are corresponding changes to the
density of states (DOS)  $n(E)$ and integrated  density of states (NOS)  $N(E)$, respectively.

$\Delta N(E)$ can be evaluated efficiently \cite{LKAG87,USPW03,EM09a} via
the so-called Lloyd formula when the underlying electronic structure is described 
by means of the multiple scattering or Korringa-Kohn-Rostoker (KKR)
formalism \cite{EKM11}. Adopting this approach we find 
%
\begin{eqnarray}
 \Delta {\cal F} &=&  -\frac{1}{\pi} \mbox{Im\, Tr\,} \int^{E_F}dE\,
                      \left(\mbox{ln}\, \underline{\underline{\tau}}(E) - \mbox{ln}\, \underline{\underline{\tau}}^{0}(E)\right) \; , 
\label{eq:Free_energy-2}
\end{eqnarray}
with the so-called scattering path operator $ \underline{\underline{\tau}}^{(0)}(E)$,
where the double underlinement indicates matrices with respect to  site and
spin-angular momentum indices \cite{EKM11}.
Within the KKR formalism these super matrices, characterizing the 
reference $\underline{\underline{\tau}}^{(0)}(E)$ and perturbed  $\underline{\underline{\tau}}(E)$
systems, respectively, are given by
%
\begin{eqnarray}
  \underline{\underline{\tau}}^{(0)}(E) &=&
          \Big[\underline{\underline{m}}^{(0)}(E) - \underline{\underline{G}}(E)\Big]^{-1}  \; ,
\label{eq:tau}
\end{eqnarray}
with   $\underline{\underline{G}}(E)$  the structure Green function and 
$\underline{\underline{m}}^{(0)}(E) =  [ \underline{\underline{t}}^{(0)}(E)  ]^{-1}$
the inverse of the corresponding site-diagonal scattering matrix that carries
all site-specific information depending on
 $\{\hat{e}_i\}$ and  $\{\vec u_i\}$
\cite{EKM11}.
 
Considering a ferromagnetic reference state ($\hat{e}_i= \hat z$)  with all atoms
in their equilibrium positions ($\vec{u}_i=0$) the perturbed state is characterized
by finite  spin tiltings $\delta \hat{e}_i$
and finite  atomic displacements of the atoms $\vec{u}_i$ for the  sites $i$.
Writing for site $i$ the resulting changes in the inverse $t$-matrix
as   $\Delta^s_{\mu} \underline{m}_i = \underline{m}_i(\delta \hat{e}_i^\mu) - \underline{m}^0_i $
and  $\Delta_{\nu}^u \underline{m}_i   = \underline{m}_i({u}_i^\nu) - \underline{m}^0_i $
allows to replace the integrand in Eq.\ (\ref{eq:Free_energy-2}) by
%
\begin{eqnarray}
  \mbox{ln} \,\underline{\underline{\tau}}
- \mbox{ln} \,\underline{\underline{\tau}}^0
  &=& - \ln \Big(1 +
       \underline{\underline{\tau}}\,[\Delta^s_{\mu} \underline{\underline{m}}_i + \Delta_{\nu}^u
                        \underline{\underline{m}}_j +  ... ] \Big) \; ,
\label{eq:tau2}
\end{eqnarray}
where all site-dependent changes in the spin configuration  $\{\hat{e}_i\}$ and atomic positions  $\{\vec u_i\}$
are accounted for in a one-to-one manner by the various terms on the right hand side. This implies in particular that the
matrices $\Delta^{s(u)}_{\mu(\nu)} \underline{\underline{m}}_i $ in Eq.\ (\ref{eq:tau2}) are site-diagonal and have non-zero blocks only for site $i$. Due to the use of the magnetic force theorem these blocks may be written 
in terms of the  spin tiltings $\delta \hat{e}_i^\mu$ and  atomic displacements of the atoms ${u}_i^\nu$ 
together with the corresponding auxiliary matrices $\underline{T}^{\mu}_i $ and ${\cal U}_{i}^{\nu}$, respectively,
as
\begin{eqnarray}
  \Delta^s_{\mu} {\underline{m}}_i & = &  \delta \hat{e}^\mu_i\,
       \underline{T}^{\mu}_i \,, \\
  \Delta^{u}_{\nu} \underline{m}_i & = & 
  u^\nu_i  \underline{\cal U}_{i}^{\nu}  \,.
                                         \label{eq:linear_distor}
\end{eqnarray}
%
Inserting these expressions into Eq.\ (\ref{eq:tau2}) and the result in turn into Eq.\ (\ref{eq:Free_energy-2})
allows us, in a straight forward way, to calculate the parameters of the
spin-lattice Hamiltonian  as the derivatives of the free energy with
respect to tilting angles and displacements. With this we derived a new
scheme to obtain systematically 
SLC terms up to any order.

In the following we will restrict ourselves to the third-order SLC 
parameters, which are linear with respect to the displacements (for more
details including the fourth-order SLC parameters see the Supplemental
Materials (SM) \cite{SuppMat}). One can write the three-site expression as:
%
\begin{eqnarray}
  {\cal J}^{\alpha\beta,\mu}_{ij,k} &=& \frac{\partial^3 {\cal F}}{\partial e^\alpha_i \,\partial e^\beta_j
                                        \, \partial u^{\mu}_k} = \frac{1}{2\pi} \mbox{Im\, Tr\,}
                                  \int^{E_F}dE \,\, \nonumber \\
   &
                          \times  & \Big[   \underline{T}^{\alpha}_i \,\underline{\tau}_{ij}
                                  \underline{T}^{\beta}_j 
                                 \, \underline{\tau}_{jk}
                                \underline{\cal U}^{\mu}_k \,\underline{\tau}_{ki}
                             +  \underline{T}^{\alpha}_i
                                   \,\underline{\tau}_{ik}
       \underline{\cal U}^{\mu}_k \,\underline{\tau}_{kj}
                                  \underline{T}^{\beta}_j 
                                 \, \underline{\tau}_{ji}\Big] \;\;
       \label{eq:Parametes_linear} 
\end{eqnarray}
We will call all these terms three-site SLC terms in the following, even if site indices are
identical. The prefactor $1/2$ occurs to avoid double counting of the identical terms upon summations in Eq.\ 
 (\ref{eq:Hamilt_extended_magneto-elastic}) over indices $i$ and $j$.

\begin{figure}[h]
\includegraphics[width=0.5\textwidth,angle=0]{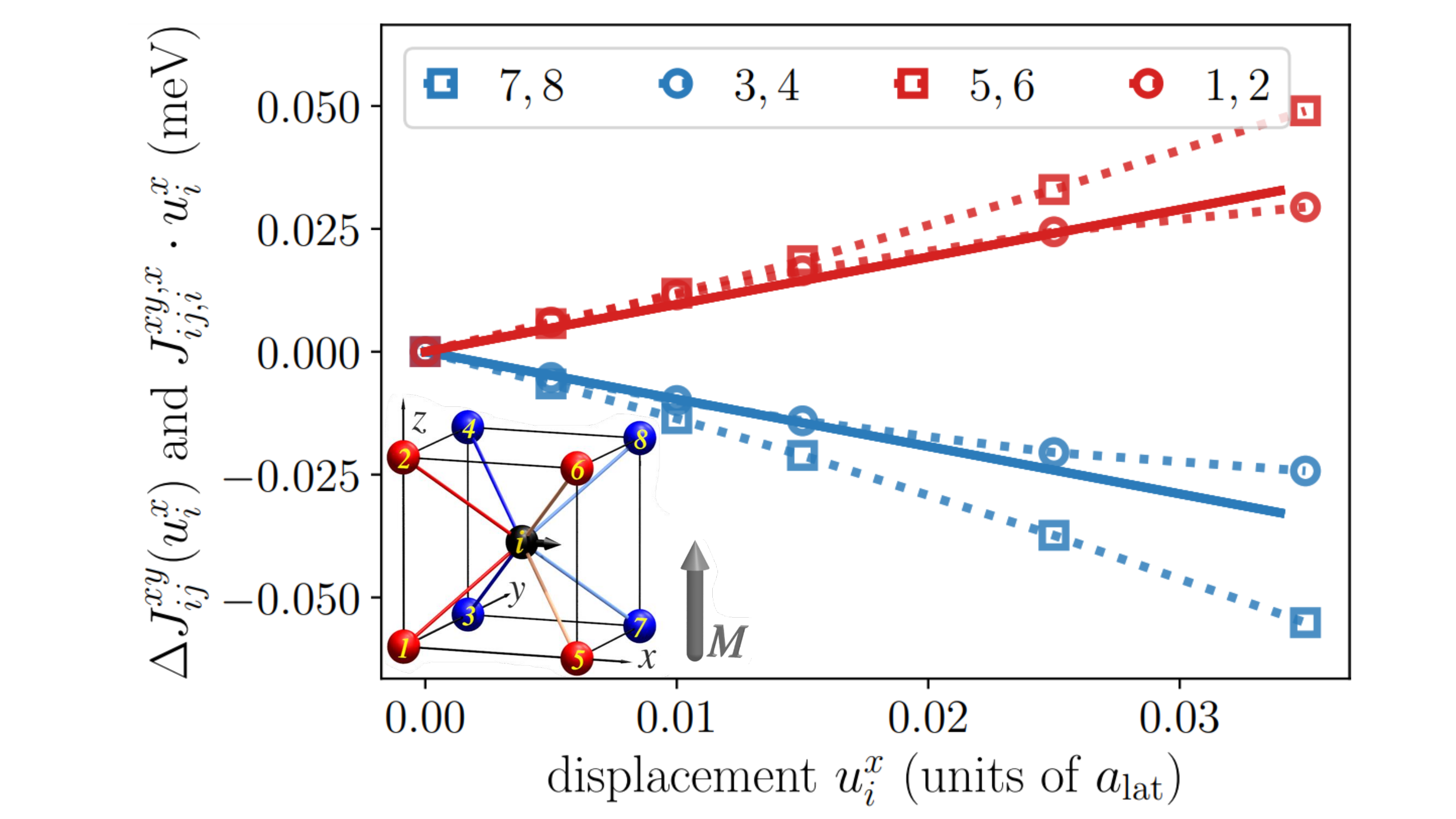}\,
\caption{\label{fig:Fe-Jij_xy-u_cmp} 
   SLC parameters versus distortion for bcc Fe: Comparison of the
   products ${\cal J}^{xy,x}_{ij,i} \cdot u^x_i$ (solid lines) with the
   corresponding SLC terms $\Delta J^{xy}_{ij}(u^x_i) = { J}^{xy}_{ij}(u^x_i)- { J}^{xy}_{ij}(0)$ calculated for a distorted system with the super cell technique (dotted lines) for one atom $i$ displaced by $u^x_i$ along the $x$-axis. The inset shows the labeling of the nearest neighbor atoms used in the figure.
}     
\end{figure}

Below, we present our first results for the SLC parameters for bcc Fe based on a ferromagnetic  
reference system with its magnetization $\vec{M}$ in
$z$-direction. Furthermore, to check the validity of our new approach
for the calculation of $ {\cal J}^{\alpha\beta,\mu}_{ij,k} $  we
performed also conventional super-cell calculations for the two-site SSC
parameters $J^{\alpha\beta}_{ij}(u^\mu_{k})$ as a function of
displacement $u^\mu_{k}$ of atom $k$. 
These calculations have been done for a  $2 \times 2 \times 2$ super cell implying a periodic displacement $u^\mu_{k}$.
For the comparison, we multiply our SLC parameters $ {\cal J}^{\alpha\beta,\mu}_{ij,k} $
with  $u^\mu_{k}$ and compare with the SSC parameters $J^{\alpha\beta}_{ij}(u^\mu_{k})$
for varying $u^\mu_{k}$, see Fig. \ref{fig:Fe-Jij_xy-u_cmp} and for a further comparison our SM \cite{SuppMat}.
If not otherwise noted, we restrict ourselves to displacments of atom $k$ along the $x$-axis, $\vec u_{k}||\hat x$. 

Since we are especially interested in the exchange of angular momentum between the spin and lattice degrees of freedom  we focus on the SOC-driven elements of the three-site SLC tensor $ {\cal J}^{\alpha\beta,\mu}_{ij,k} $, which give rise to magneto-crystalline anisotropies (MCA) and Dzyaloshinskii-Moriya interactions (DMI). 
The interatomic three-site SLC parameters  
represent a 'non-local' contribution to the MCA \cite{USPW03} as well as DMI induced by a
displacement of atom $k$. Fig.\ \ref{fig:Fe-Jij_xy-u_cmp} shows, as an
example, the nearest-neighbor SLC products ${\cal J}^{xy,x}_{ij,i} \cdot
u^x_i$ compared with the two-site SSC interaction  parameters for four
groups of atoms as sketched in the inset of Fig.\
\ref{fig:Fe-Jij_xy-u_cmp}. Obviously, one finds good agreement between
the different approaches for small amplitudes of the displacement
$u^x_i$. Here one should stress that the present
approach gives direct access to the SLC parameters
of any order with respect to the atomic displacement
or spin tilting in an extremely efficient way.


  The anti-symmetric part of the off-diagonal SLC parameters, ${\cal J}^{{\rm
    off-a},\mu}_{ij,k}  =  \frac{1}{2}({\cal J}^{xy,\mu}_{ij,k} - {\cal J}^{yx,\mu}_{ij,k})$, can be interpreted
\cite{USPW03} as the DMI, $D^z_{ij}$, induced by the symmetry-breaking displacement 
of atom $k$ and one can define a Dzyaloshinskii-Moriya-like spin lattice coupling (DSLC), ${\cal D}^{z,\mu}_{ij,k} = {\cal J}^{{\rm off-a},\mu}_{ij,k}$. Note that the conventional DMI vanishes for the non-distorted bcc Fe lattice due to inversion symmetry. Furthermore,  for symmetry reasons, all anti-symmetric off-diagonal elements of the three-site SLC are equal 
to zero in the case of a displacement of atom $k$, positioned at the same distance from atoms $i$ and $j$, along $\hat{z}$, implying ${\cal D}^{z,z}_{ij,k} = 0$. This does not apply for the other components, ${\cal D}^{x,z}_{ij,k}$ and
${\cal D}^{y,z}_{ij,k}$.

\begin{figure}[h]
\includegraphics[width=0.44\textwidth,angle=0]{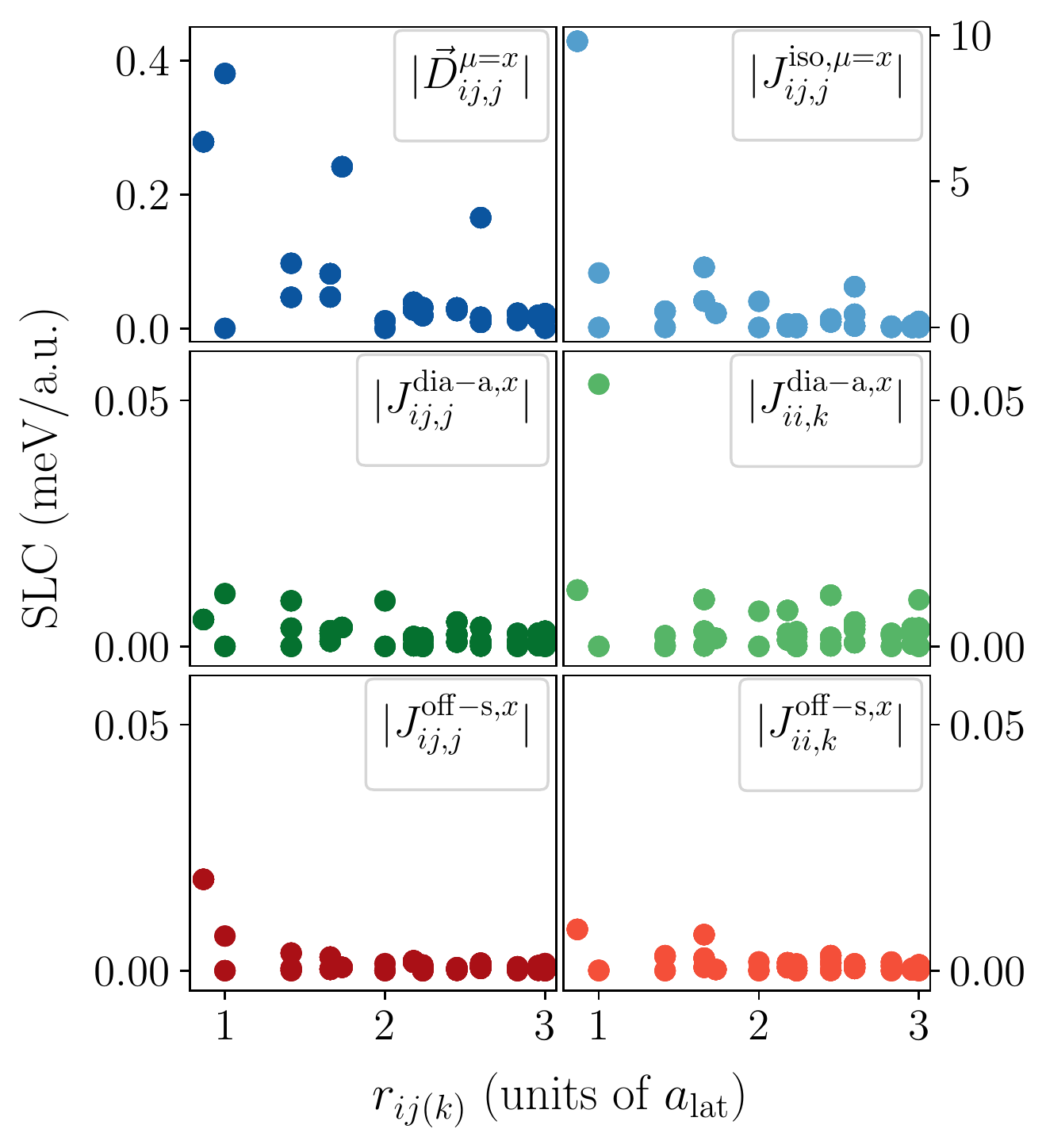}\,
\caption{\label{fig:DSLC_vs_Rij}  
  Magnitude of site-off-diagonal and site-diagonal SLC parameters: DMI $|\vec{\cal D}^x_{ij,j}|$ and isotropic SLC ${\cal J}^{\mathrm{iso},x}_{ij,j}$ (top), anti-symmetric diagonal components ${\cal J}^{\mathrm{dia-a},x}_{ij,j}$ and ${\cal J}^{\mathrm{dia-a},x}_{ii,k}$ (middle), and symmetric off-diagonal components ${\cal J}^{\mathrm{off-s},x}_{ij,j}$ and ${\cal J}^{\mathrm{off-s},x}_{ii,k}$ (bottom).
}     
\end{figure}

A further analysis of our SLC parameters in this context is shown in Fig. \ref{fig:DSLC_vs_Rij}, again for $i = k$ which implies that the displacement along $x$ direction is applied to one of the interacting atoms. Results for $k \neq j(i)$
are shown in the SM \cite{SuppMat}.  Different components of the SCL parameters are plotted as a function of the
distance $r_{ij}$. The absolute values of the DSLC parameters $|\vec{\cal D}|^{\mu=x}_{ij,k}$ show a rather slow decay 
with the distance $r_{ij}$.  These parameters can take different values for the same distance, which is a
result of the symmetry imposed vanishing of certain components of the
DMI-like SLC for some $\vec{r}_{ij}$ directions, which depend in turn also on 
the direction of the displacement $\vec{u}$. The isotropic SLC parameters  ${\cal
  J}^{\mathrm{iso},\mu=x}_{ij,j}$, which have only a weak dependence on the SOC, are about one order of magnitude larger
than the DSLC. All other SOC-driven parameters shown in Fig. \ref{fig:DSLC_vs_Rij}, characterizing the
displacement-induced contributions to MCA, are much smaller than the DSLC.

  Calculating the  ${\cal D}^{\nu,\mu}_{ij,k}$ parameters for ordered FePt in the CuAu structure, with Fe atoms
  on sites $i$ and $j$, we demonstrate in the SM \cite{SuppMat} a key role of the SOC on the atoms mediating these interactions. It follows that the DSLC parameters  ${\cal D}^{\nu,\mu}_{ij,k}$ (both, for $k = j$ and $k \neq j$) exhibit only
  weak dependence on SOC strength of Fe sites, while they
  decrease by about an order of magnitude when the SOC on Pt sites goes to
  zero, indicating the key role of the atoms mediating the coupling between the atoms $i$ and $j$ for the DSLC magnitude.
A more detailed analysis of the properties of three-site SLC parameters  is
given in the SM \cite{SuppMat}. The role of the
dipolar contribution to the SLC is not discussed here
as it is expected to play a minor role for bcc Fe.

\begin{figure}[h]
\includegraphics[width=0.36\textwidth,angle=0,clip]{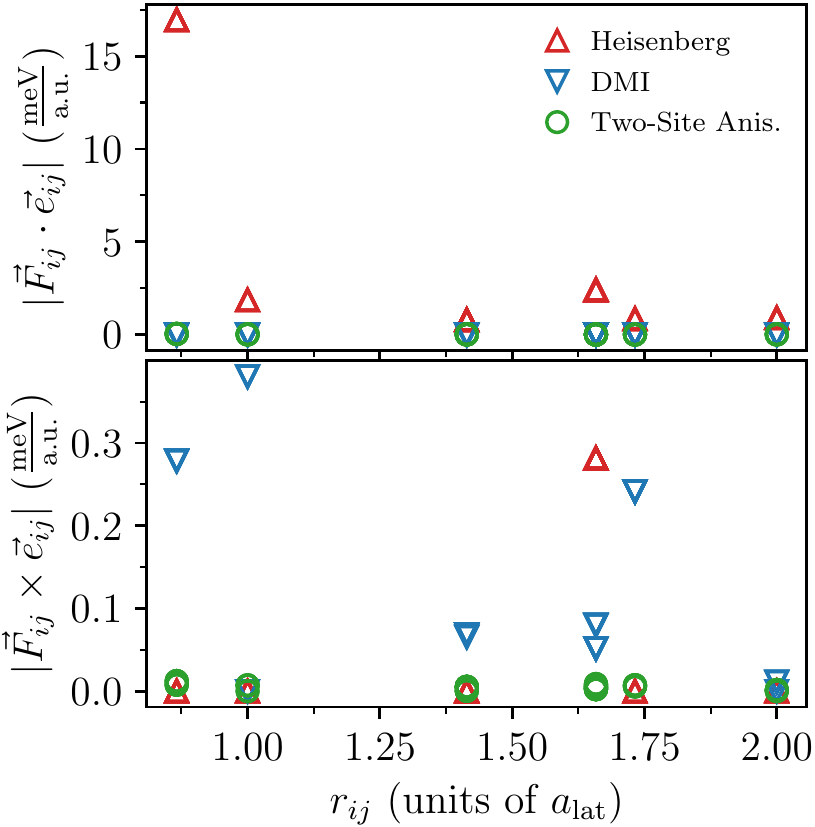}\,a)
\includegraphics[width=0.18\textwidth,angle=0,clip]{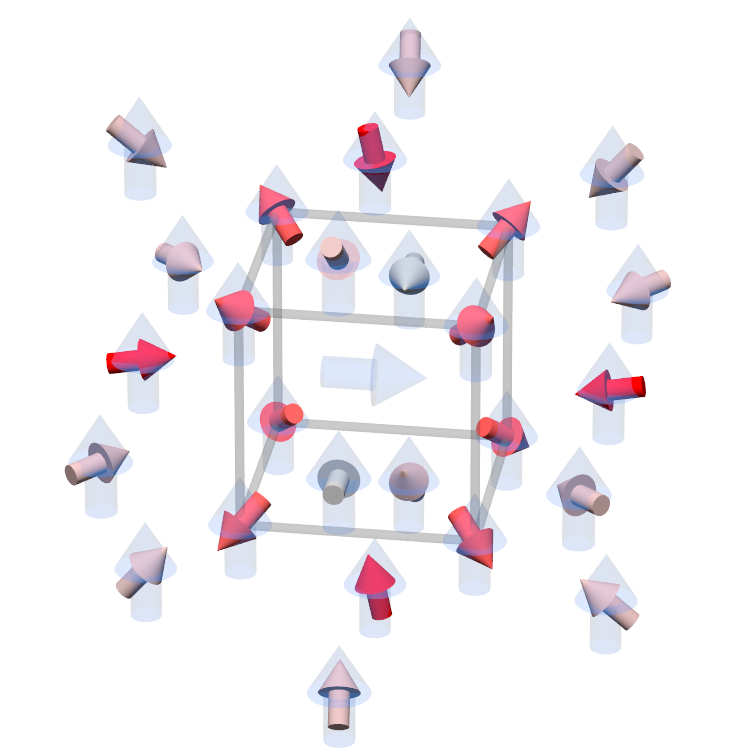}\,b)
\includegraphics[width=0.16\textwidth,angle=0,clip]{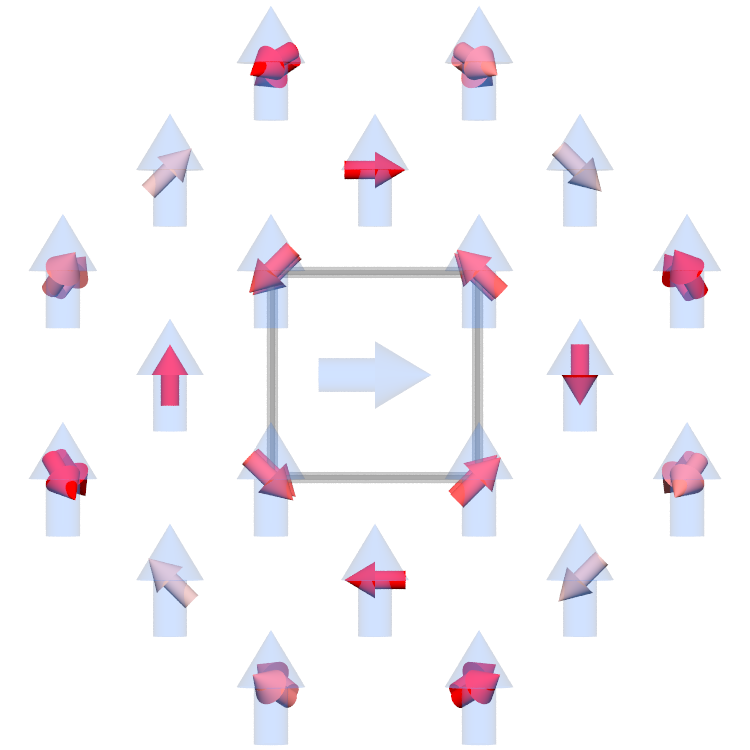}\, c)
\caption{\label{fig:Forces}
a) Longitudinal (top) and perpendicular (bottom) forces (with respect to
$\vec{r}_{ij}$) emerging from the isotropic, DMI-type and
MCA-type contributions to the SLC parameters. The central spin at $\vec{r}=(0,0,0)$ is tilted in
$y$-direction while all the others point in $z$-direction (transparent arrows in b) and c)).
b) Directions of the forces with color coding indicating their perpendicular components with red
being high and grey being low. c) Directions of the perpendicular forces with the same color coding as in b). 
}
\end{figure}
In an attempt to estimate the forces that act on the atoms for a
non-collinear spin configuration we assume naively that these are given
via the gradient of our spin-lattice Hamiltonian, Eq. (\ref{eq:Hamilt_extended_magneto-elastic}), with respect to the
displacements, $\vec{F} = - \frac{\partial {\cal H}_{\rm{sl}}}{\partial \vec{u}_k}$, see also \cite{SHNE08}. We note, however, that this spin-lattice Hamiltonian does not conserve the total (spin plus lattice) angular momentum which
questions this approach. However, within this approach we are able to
calculate the forces   acting on neighboring atoms of a
central spin which is tilted in $y$-direction while all the other spins are aligned in
$x$-direction. Because of this non-collinear spin configuration forces
act on the neighboring atoms which are shown in Fig. \ref{fig:Forces}. These forces are decomposed into their longitudinal and perpendicular parts (with respect to the distance vector $\vec{r}_{ij}$)
and whether they come from isotropic, DMI-type or MCA-type contributions to the SLC parameters. 
The main contribution to the longitudinal forces, which do not transfer
angular momentum, comes once again from the isotropic Heisenberg
interaction. However, the largest perpendicular --- and with that
angular momentum transferring --- component is once again due to the
DMI-type interaction. Plugging in numbers for the mass of an Fe atom we
note that the latter is sufficient to accelerate a free Fe atom within
100 fs to a velocitiy of about 1 m/s. 



The role of the DMI-like SLC parameters for the spin-lattice
angular momentum transfer can also be discussed in terms of
magnon-phonon interactions \cite{RKB+14,GC15,SVSB19,RSBD20}.
With this in mind, we represent the second term of the 
Hamiltonian in Eq.\ (\ref{eq:Hamilt_extended_magneto-elastic}), which is
linear with respect to displacement $\vec{u}_k$, in terms of isotropic
and anisotropic interactions focusing on the DSLC contribution
\begin{eqnarray}
  {\cal H}_{\rm{me-DMI}} &=&\frac{1}{S^2} \sum_{i,j} \sum_{k,\mu}
                          \vec{\cal D}_{ij,k}^{\mu} \cdot (\hat{s}_i \times \hat{s}_j)
                          u^{\mu}_k  \,.
\label{eq:Hamilt_exchange_magneto-elastic-linear-3}
\end{eqnarray}
${\cal H}_{\rm{me-DMI}}$ is now given in a form using spin operators $\hat{s}^\alpha_i$
instead of spin orientation vectors $\hat{e}^\alpha_i$, where $S$ is the 
maximal value for spin moment per atom. 
For the sake of simplicity, we consider again a ferromagnetic system with one atom per unit cell as, e.g., bcc Fe, and with the magnetization $M$ along $z$ direction. To describe the magnon-phonon
interactions, it is customary to introduce spin raising and lowering
operators $\hat{s}^\pm_i = \hat{s}^x_i \pm i\hat{s}^y_i$.
Using the Holstein-Primakoff transformation \cite{HP40,SuppMat}, 
local spin fluctuations can be represented in terms of canonical
boson operators, $\hat{b}^\dagger_i$ and $\hat{b}_i$.
Performing Fourier transformations for the spin operators and the
atomic displacements the Hamiltonian ${\cal
  H}_{\rm{me-DMI}}$ can be reduced to the momentum representation form \cite{SuppMat} 
\begin{eqnarray}
{\cal H}_{\rm{me-DMI}}   &=& \frac{2i}{\sqrt{2S}}  \sum_{\vec{q},\mu} 
                         \bigg[ {\cal D}_{\vec{q}}^{-,\mu} 
                           \hat{b}_{\vec{q}}       
                          - {\cal D}_{-\vec{q}}^{+,\mu} 
                           \hat{b}^\dagger_{-\vec{q}}                         
                             \bigg] {u}^{\mu}_{\vec{q}} \; \nonumber \\
 &&- \; \frac{2i}{\sqrt{N}}
      \frac{1}{S}\sum_{\vec{k}, \vec{k}', \mu}
                         {\cal D}_{\vec{k},\vec{k}'}^{z,\mu} 
                             \, \hat{b}^\dagger_{\vec{k}}
                             \hat{b}_{\vec{k}'}\, {u}^{\mu}_{(\vec{k}' - \vec{k})}      \,,
\label{eq:Hamilt_exchange_magneto-elastic-linear-5}
\end{eqnarray}
where ${u}^{\mu}_{\vec{q}} = \sum_{\lambda}
\epsilon^{\mu}_{\lambda,\vec{q}} X_{\lambda,\vec{q}}$ is  
the eigenvector corresponding to the phonon mode $(\lambda,\vec{q})$ with
 the polarization $\epsilon^{\mu}_{\lambda,\vec{q}}$. In the
second quantization form for the phonon modes $X_{\lambda,\vec{q}}$ can
be given in terms of the phonon creation and annihilation operators,
respectively \cite{Boe83,SuppMat}. Following the conclusions in Ref.\
\cite{RSBD20}, the first term in Eq.\ (\ref{eq:Hamilt_exchange_magneto-elastic-linear-5}), which is
determined by the DSLC components ${\cal D}_{ij,k}^{x,\mu}$ and ${\cal D}_{ij,k}^{y,\mu}$,
describes the magnon-phonon scattering that allows for an exchange of angular momentum.  
On the other hand, the  DSLC component ${\cal D}_{ij,k}^{z,\mu}$ (with $z$ the magnetization direction)
contributes to the magnon-number conserving scattering characterized by
the energy transfer only (see Ref.\ \onlinecite{RSBD20}). 

\begin{figure}[h]
\includegraphics[width=0.52\textwidth,angle=0,clip]{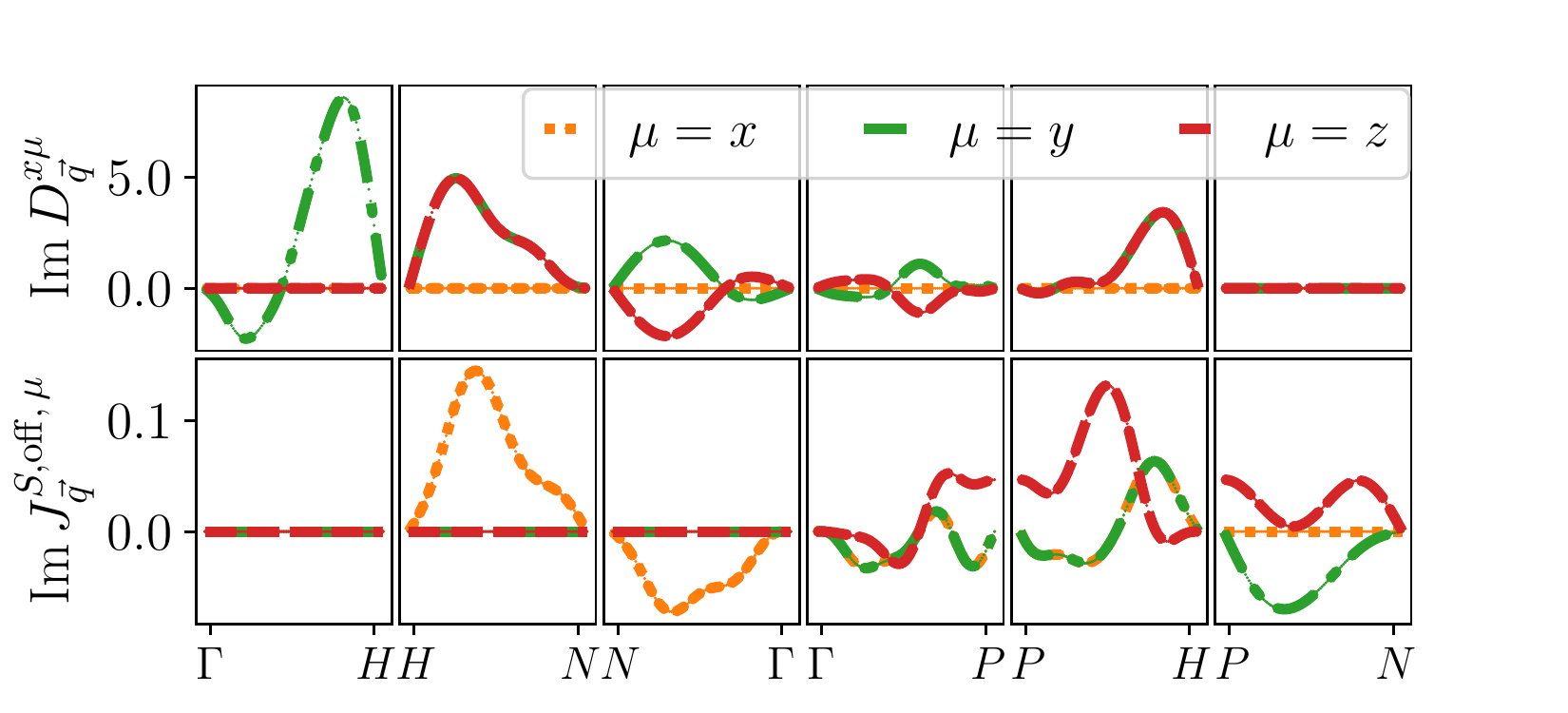}\,
\caption{\label{fig:SLC-q_1} Imaginary part $\mbox{Im}\,{\cal D}_{\vec{q}}^{x,\mu}$
  (top) and $\mathrm{Im}(J_{\vec{q}}^{xy,\mu} + J_{\vec{q}}^{yx,\mu})/2$
  (bottom) of the SLC parameters for bcc Fe, with $\mu = x,y,z$, 
  plotted for $\vec{q}$ along high-symmetry lines of the 
 Brillouin zone. The real part is zero in all cases.   }     
\end{figure}

The upper part of Fig. \ref{fig:SLC-q_1} shows the Fourier transforms 
${\cal D}_{\vec{q}}^{x,\mu}$ of the SLC parameters for bcc Fe, given by the expression:
\begin{eqnarray}
  {\cal D}_{\vec{q}}^{\nu,\mu} &=& \sum_{j,k} {\cal D}_{ij,k}^{\nu,\mu}
                            e^{i\vec{q}\cdot(\vec{R}_{j} - \vec{R}_{i})}
                            e^{-i\vec{q}\cdot(\vec{R}_{k} - \vec{R}_{i})}  \;,
\label{eq:DMI_xy_Fourier}
\end{eqnarray}
which are plotted for $\vec{q}$ along the high-symmetry lines in
the Brillouin zone (see also SM \cite{SuppMat}). One finds a strong dependence on the direction 
and absolute value of wave vector $\vec{q}$. Moreover, this dependence is different for different components of the atomic displacements, which reflects in turn the difference of the magnon-phonon interactions for the
transverse and longitudinal phonon modes.
In addition, the lower part of Fig. \ref{fig:SLC-q_1} shows the 
Fourier transforms of the cite-diagonal SLC parameters
$\mathrm{Im}(J_{\vec{q}}^{xy,\mu} + J_{\vec{q}}^{yx,\mu})/2$, with $\mu =
x,y,z$, which can be seen as a local contribution to the magnetic anisotropy induced by atomic displacements. Comparing the results confirm the dominating role of the DSLC for the magnon-phonon 
hybridization.


The contribution of the first term in the DSLC Hamiltonian ${\cal H}_{\rm{me-DMI}}$ in Eq.\
(\ref{eq:Hamilt_exchange_magneto-elastic-linear-5}) to the phonon angular
momentum dynamics can also be characterized by the corresponding torque on the
lattice produced by the spin system \cite{RSBD20}.
In particular, the torque on the phonon spin entering the equation of
motion for the phonon angular momentum 
$\vec{\cal L}_{ph} = \sum_{k} \vec{u}_k \times \vec{\pi}_k$ is given
by the expression \cite{RSBD20} 
\begin{eqnarray}
  \vec{\cal T}_{\rm{me}} &=& -\sum_{k} \vec{u}_k \times \frac{\partial {\cal
                          H}_{\rm{me}}}{\partial \vec{u}_k}
\label{eq:Ang_mom_2}
\end{eqnarray}
with the $\vec{u}_i$ and $\vec{\pi}_i$ the displacement and linear
momentum operators, respectively.

Thus, the one magnon - one phonon contribution to the torque, $\vec{\cal
  T}_{\rm{me}}$, originating from the first term of the DSLC Hamiltonian
${\cal H}_{\rm{me-DMI}}$ in  Eq.\
(\ref{eq:Hamilt_exchange_magneto-elastic-linear-5}), is  
given by the expression (see SM \cite{SuppMat})
\begin{eqnarray}
  {\cal T}^\gamma_{\rm{me-DMI}} &=& \frac{2i}{\sqrt{2S}}   \sum_{\vec{q},\lambda} 
                        \bigg[  {\Gamma}_{\lambda,\vec{q}}^{-,\gamma}     
    \hat{b}_{\vec{q}}    
    -  {\Gamma}_{\lambda,-\vec{q}}^{+,\gamma} \hat{b}^\dagger_{-\vec{q}} 
    \bigg]  X_{\lambda,\vec{q}}                              \,,  
\label{eq:Hamilt_DMI_magneto-elastic-linear-3}
\end{eqnarray}
Here, the interaction vertices ${\Gamma}_{\lambda,\vec{q}}^{\pm,\gamma}$ are defined as 
\begin{eqnarray}
  {\Gamma}_{\lambda,\vec{q}}^{\pm,\gamma} &=& \epsilon_{\alpha\beta\gamma}\bigg[ \epsilon^{\alpha}_{\lambda,\vec{q}} 
                      {\cal D}_{\vec{q}}^{\pm,\beta} -\epsilon^{\beta}_{\lambda,\vec{q}}  {\cal D}_{\vec{q}}^{\pm,\alpha}\bigg] \,.
\end{eqnarray}
with $\epsilon_{\alpha\beta\gamma}$ the Levi-Civita symbol. As one can
see they are fully determined by the DSLC parameters discussed above.


In summary, we present a scheme to calculate microscopic and relativistic spin-lattice coupling
parameters from first-principles. The perturbation due to a lattice
distortion is treated on the same footing as the distortion due to spin
tilting, giving  access to SLC parameters up to any order for these
perturbations.  
In linear order of the displacements our parameters can be compared with
spin-spin coupling 
parameters obtained via super-cell calculations 
with an applied distortion \cite{HTM+19} and good agreement is found. 
Analyzing the properties of these parameters, in particular their
dependence on spin-orbit coupling,  
we find that even in bcc Fe the leading term which is responsibla for
the exchange of angular momentum between the spin system and the lattice
is a Dzyaloshiskii-Moriya-type interaction, which emerges due to the
symmetry breaking distortion of the lattice. Our findings, hence, stress
the importance of relativistic effects for the transfer of angular
momentum from magnonic excitations to circularily polarized phonons
\cite{Tau22,GC15,MKL19,SVSB19,RSBD20}

\section*{Acknowledgements}


The work in Konstanz was supported by the DFG via SFB 1432 and Project
No. NO 290/5-1.


\begin{thebibliography}{28}%
\makeatletter
\providecommand \@ifxundefined [1]{%
 \@ifx{#1\undefined}
}%
\providecommand \@ifnum [1]{%
 \ifnum #1\expandafter \@firstoftwo
 \else \expandafter \@secondoftwo
 \fi
}%
\providecommand \@ifx [1]{%
 \ifx #1\expandafter \@firstoftwo
 \else \expandafter \@secondoftwo
 \fi
}%
\providecommand \natexlab [1]{#1}%
\providecommand \enquote  [1]{``#1''}%
\providecommand \bibnamefont  [1]{#1}%
\providecommand \bibfnamefont [1]{#1}%
\providecommand \citenamefont [1]{#1}%
\providecommand \href@noop [0]{\@secondoftwo}%
\providecommand \href [0]{\begingroup \@sanitize@url \@href}%
\providecommand \@href[1]{\@@startlink{#1}\@@href}%
\providecommand \@@href[1]{\endgroup#1\@@endlink}%
\providecommand \@sanitize@url [0]{\catcode `\\12\catcode `\$12\catcode
  `\&12\catcode `\#12\catcode `\^12\catcode `\_12\catcode `\%12\relax}%
\providecommand \@@startlink[1]{}%
\providecommand \@@endlink[0]{}%
\providecommand \url  [0]{\begingroup\@sanitize@url \@url }%
\providecommand \@url [1]{\endgroup\@href {#1}{\urlprefix }}%
\providecommand \urlprefix  [0]{URL }%
\providecommand \Eprint [0]{\href }%
\providecommand \doibase [0]{http://dx.doi.org/}%
\providecommand \selectlanguage [0]{\@gobble}%
\providecommand \bibinfo  [0]{\@secondoftwo}%
\providecommand \bibfield  [0]{\@secondoftwo}%
\providecommand \translation [1]{[#1]}%
\providecommand \BibitemOpen [0]{}%
\providecommand \bibitemStop [0]{}%
\providecommand \bibitemNoStop [0]{.\EOS\space}%
\providecommand \EOS [0]{\spacefactor3000\relax}%
\providecommand \BibitemShut  [1]{\csname bibitem#1\endcsname}%
\let\auto@bib@innerbib\@empty
\bibitem [{\citenamefont {Hirohata}\ \emph {et~al.}(2020)\citenamefont
  {Hirohata}, \citenamefont {Yamada}, \citenamefont {Nakatani}, \citenamefont
  {Prejbeanu}, \citenamefont {Diény}, \citenamefont {Pirro},\ and\
  \citenamefont {Hillebrands}}]{Hir20}%
  \BibitemOpen
  \bibfield  {author} {\bibinfo {author} {\bibfnamefont {A.}~\bibnamefont
  {Hirohata}}, \bibinfo {author} {\bibfnamefont {K.}~\bibnamefont {Yamada}},
  \bibinfo {author} {\bibfnamefont {Y.}~\bibnamefont {Nakatani}}, \bibinfo
  {author} {\bibfnamefont {I.-L.}\ \bibnamefont {Prejbeanu}}, \bibinfo {author}
  {\bibfnamefont {B.}~\bibnamefont {Diény}}, \bibinfo {author} {\bibfnamefont
  {P.}~\bibnamefont {Pirro}}, \ and\ \bibinfo {author} {\bibfnamefont
  {B.}~\bibnamefont {Hillebrands}},\ }\href {\doibase
  https://doi.org/10.1016/j.jmmm.2020.166711} {\bibfield  {journal} {\bibinfo
  {journal} {Journal of Magnetism and Magnetic Materials}\ }\textbf {\bibinfo
  {volume} {509}},\ \bibinfo {pages} {166711} (\bibinfo {year}
  {2020})}\BibitemShut {NoStop}%
\bibitem [{\citenamefont {Radu}\ \emph {et~al.}(2011)\citenamefont {Radu},
  \citenamefont {Vahaplar}, \citenamefont {Stamm}, \citenamefont {Kachel},
  \citenamefont {Pontius}, \citenamefont {D\"urr}, \citenamefont {Ostler},
  \citenamefont {Evans}, \citenamefont {Chantrell}, \citenamefont {Tsukamoto},
  \citenamefont {Itoh}, \citenamefont {Kirilyuk}, \citenamefont {Rasing},\ and\
  \citenamefont {Kimel}}]{Rad11}%
  \BibitemOpen
  \bibfield  {author} {\bibinfo {author} {\bibfnamefont {I.}~\bibnamefont
  {Radu}}, \bibinfo {author} {\bibfnamefont {K.}~\bibnamefont {Vahaplar}},
  \bibinfo {author} {\bibfnamefont {C.}~\bibnamefont {Stamm}}, \bibinfo
  {author} {\bibfnamefont {T.}~\bibnamefont {Kachel}}, \bibinfo {author}
  {\bibfnamefont {N.}~\bibnamefont {Pontius}}, \bibinfo {author} {\bibfnamefont
  {H.~A.}\ \bibnamefont {D\"urr}}, \bibinfo {author} {\bibfnamefont
  {J.}~\bibnamefont {Ostler}, \bibfnamefont {T.~A. abd~Barker}}, \bibinfo
  {author} {\bibfnamefont {R.~F.~L.}\ \bibnamefont {Evans}}, \bibinfo {author}
  {\bibfnamefont {R.~W.}\ \bibnamefont {Chantrell}}, \bibinfo {author}
  {\bibfnamefont {A.}~\bibnamefont {Tsukamoto}}, \bibinfo {author}
  {\bibfnamefont {A.}~\bibnamefont {Itoh}}, \bibinfo {author} {\bibfnamefont
  {A.}~\bibnamefont {Kirilyuk}}, \bibinfo {author} {\bibfnamefont
  {T.}~\bibnamefont {Rasing}}, \ and\ \bibinfo {author} {\bibfnamefont {A.~V.}\
  \bibnamefont {Kimel}},\ }\href {\doibase https://doi.org/10.1038/nature09901}
  {\bibfield  {journal} {\bibinfo  {journal} {Nature}\ }\textbf {\bibinfo
  {volume} {472}},\ \bibinfo {pages} {205} (\bibinfo {year}
  {2011})}\BibitemShut {NoStop}%
\bibitem [{\citenamefont {Mentink}\ \emph {et~al.}(2012)\citenamefont
  {Mentink}, \citenamefont {Hellsvik}, \citenamefont {Afanasiev}, \citenamefont
  {Ivanov}, \citenamefont {Kirilyuk}, \citenamefont {Kimel}, \citenamefont
  {Eriksson}, \citenamefont {Katsnelson},\ and\ \citenamefont
  {Rasing}}]{Men12}%
  \BibitemOpen
  \bibfield  {author} {\bibinfo {author} {\bibfnamefont {J.~H.}\ \bibnamefont
  {Mentink}}, \bibinfo {author} {\bibfnamefont {J.}~\bibnamefont {Hellsvik}},
  \bibinfo {author} {\bibfnamefont {D.~V.}\ \bibnamefont {Afanasiev}}, \bibinfo
  {author} {\bibfnamefont {B.~A.}\ \bibnamefont {Ivanov}}, \bibinfo {author}
  {\bibfnamefont {A.}~\bibnamefont {Kirilyuk}}, \bibinfo {author}
  {\bibfnamefont {A.~V.}\ \bibnamefont {Kimel}}, \bibinfo {author}
  {\bibfnamefont {O.}~\bibnamefont {Eriksson}}, \bibinfo {author}
  {\bibfnamefont {M.~I.}\ \bibnamefont {Katsnelson}}, \ and\ \bibinfo {author}
  {\bibfnamefont {T.}~\bibnamefont {Rasing}},\ }\href {\doibase
  10.1103/PhysRevLett.108.057202} {\bibfield  {journal} {\bibinfo  {journal}
  {Phys. Rev. Lett.}\ }\textbf {\bibinfo {volume} {108}},\ \bibinfo {pages}
  {057202} (\bibinfo {year} {2012})}\BibitemShut {NoStop}%
\bibitem [{\citenamefont {Wienholdt}\ \emph {et~al.}(2013)\citenamefont
  {Wienholdt}, \citenamefont {Hinzke}, \citenamefont {Carva}, \citenamefont
  {Oppeneer},\ and\ \citenamefont {Nowak}}]{WHC+13}%
  \BibitemOpen
  \bibfield  {author} {\bibinfo {author} {\bibfnamefont {S.}~\bibnamefont
  {Wienholdt}}, \bibinfo {author} {\bibfnamefont {D.}~\bibnamefont {Hinzke}},
  \bibinfo {author} {\bibfnamefont {K.}~\bibnamefont {Carva}}, \bibinfo
  {author} {\bibfnamefont {P.~M.}\ \bibnamefont {Oppeneer}}, \ and\ \bibinfo
  {author} {\bibfnamefont {U.}~\bibnamefont {Nowak}},\ }\href {\doibase
  10.1103/PhysRevB.88.020406} {\bibfield  {journal} {\bibinfo  {journal} {Phys.
  Rev. B}\ }\textbf {\bibinfo {volume} {88}},\ \bibinfo {pages} {020406}
  (\bibinfo {year} {2013})}\BibitemShut {NoStop}%
\bibitem [{\citenamefont {Tauchert}\ \emph {et~al.}(2022)\citenamefont
  {Tauchert}, \citenamefont {Volkov}, \citenamefont {Ehberger}, \citenamefont
  {Kazenwadel}, \citenamefont {Evers}, \citenamefont {Lange}, \citenamefont
  {Donges}, \citenamefont {Book}, \citenamefont {Kreuzpaintner}, \citenamefont
  {Nowak},\ and\ \citenamefont {Baum}}]{Tau22}%
  \BibitemOpen
  \bibfield  {author} {\bibinfo {author} {\bibfnamefont {S.~R.}\ \bibnamefont
  {Tauchert}}, \bibinfo {author} {\bibfnamefont {M.}~\bibnamefont {Volkov}},
  \bibinfo {author} {\bibfnamefont {D.}~\bibnamefont {Ehberger}}, \bibinfo
  {author} {\bibfnamefont {D.}~\bibnamefont {Kazenwadel}}, \bibinfo {author}
  {\bibfnamefont {M.}~\bibnamefont {Evers}}, \bibinfo {author} {\bibfnamefont
  {H.}~\bibnamefont {Lange}}, \bibinfo {author} {\bibfnamefont
  {A.}~\bibnamefont {Donges}}, \bibinfo {author} {\bibfnamefont
  {A.}~\bibnamefont {Book}}, \bibinfo {author} {\bibfnamefont {W.}~\bibnamefont
  {Kreuzpaintner}}, \bibinfo {author} {\bibfnamefont {U.}~\bibnamefont
  {Nowak}}, \ and\ \bibinfo {author} {\bibfnamefont {P.}~\bibnamefont {Baum}},\
  }\href {\doibase https://doi.org/10.1038/s41586-021-04306-4} {\bibfield
  {journal} {\bibinfo  {journal} {Nature}\ }\textbf {\bibinfo {volume} {602}},\
  \bibinfo {pages} {73} (\bibinfo {year} {2022})}\BibitemShut {NoStop}%
\bibitem [{\citenamefont {Dornes}\ \emph {et~al.}(2019)\citenamefont {Dornes},
  \citenamefont {Acremann}, \citenamefont {Savoini}, \citenamefont {Kubli},
  \citenamefont {Neugebauer}, \citenamefont {Abreu}, \citenamefont {Huber},
  \citenamefont {Lantz}, \citenamefont {Vaz}, \citenamefont {Lemke},
  \citenamefont {Bothschafter}, \citenamefont {Porer}, \citenamefont
  {Esposito}, \citenamefont {Rettig}, \citenamefont {Buzzi}, \citenamefont
  {Alberca}, \citenamefont {Windsor}, \citenamefont {Beaud}, \citenamefont
  {Staub}, \citenamefont {Zhu}, \citenamefont {Song}, \citenamefont {Glownia},\
  and\ \citenamefont {Johnson}}]{DAS+19}%
  \BibitemOpen
  \bibfield  {author} {\bibinfo {author} {\bibfnamefont {C.}~\bibnamefont
  {Dornes}}, \bibinfo {author} {\bibfnamefont {Y.}~\bibnamefont {Acremann}},
  \bibinfo {author} {\bibfnamefont {M.}~\bibnamefont {Savoini}}, \bibinfo
  {author} {\bibfnamefont {M.}~\bibnamefont {Kubli}}, \bibinfo {author}
  {\bibfnamefont {M.~J.}\ \bibnamefont {Neugebauer}}, \bibinfo {author}
  {\bibfnamefont {E.}~\bibnamefont {Abreu}}, \bibinfo {author} {\bibfnamefont
  {L.}~\bibnamefont {Huber}}, \bibinfo {author} {\bibfnamefont
  {G.}~\bibnamefont {Lantz}}, \bibinfo {author} {\bibfnamefont {C.~A.~F.}\
  \bibnamefont {Vaz}}, \bibinfo {author} {\bibfnamefont {H.}~\bibnamefont
  {Lemke}}, \bibinfo {author} {\bibfnamefont {E.~M.}\ \bibnamefont
  {Bothschafter}}, \bibinfo {author} {\bibfnamefont {M.}~\bibnamefont {Porer}},
  \bibinfo {author} {\bibfnamefont {V.}~\bibnamefont {Esposito}}, \bibinfo
  {author} {\bibfnamefont {L.}~\bibnamefont {Rettig}}, \bibinfo {author}
  {\bibfnamefont {M.}~\bibnamefont {Buzzi}}, \bibinfo {author} {\bibfnamefont
  {A.}~\bibnamefont {Alberca}}, \bibinfo {author} {\bibfnamefont {Y.~W.}\
  \bibnamefont {Windsor}}, \bibinfo {author} {\bibfnamefont {P.}~\bibnamefont
  {Beaud}}, \bibinfo {author} {\bibfnamefont {U.}~\bibnamefont {Staub}},
  \bibinfo {author} {\bibfnamefont {D.}~\bibnamefont {Zhu}}, \bibinfo {author}
  {\bibfnamefont {S.}~\bibnamefont {Song}}, \bibinfo {author} {\bibfnamefont
  {J.~M.}\ \bibnamefont {Glownia}}, \ and\ \bibinfo {author} {\bibfnamefont
  {S.~L.}\ \bibnamefont {Johnson}},\ }\href
  {https://doi.org/10.1038/s41586-018-0822-7} {\bibfield  {journal} {\bibinfo
  {journal} {Nature}\ }\textbf {\bibinfo {volume} {565}},\ \bibinfo {pages}
  {209} (\bibinfo {year} {2019})}\BibitemShut {NoStop}%
\bibitem [{\citenamefont {Garanin}\ and\ \citenamefont
  {Chudnovsky}(2015)}]{GC15}%
  \BibitemOpen
  \bibfield  {author} {\bibinfo {author} {\bibfnamefont {D.~A.}\ \bibnamefont
  {Garanin}}\ and\ \bibinfo {author} {\bibfnamefont {E.~M.}\ \bibnamefont
  {Chudnovsky}},\ }\href {\doibase 10.1103/PhysRevB.92.024421} {\bibfield
  {journal} {\bibinfo  {journal} {Phys. Rev. B}\ }\textbf {\bibinfo {volume}
  {92}},\ \bibinfo {pages} {024421} (\bibinfo {year} {2015})}\BibitemShut
  {NoStop}%
\bibitem [{\citenamefont {Mentink}\ \emph {et~al.}(2019)\citenamefont
  {Mentink}, \citenamefont {Katsnelson},\ and\ \citenamefont
  {Lemeshko}}]{MKL19}%
  \BibitemOpen
  \bibfield  {author} {\bibinfo {author} {\bibfnamefont {J.~H.}\ \bibnamefont
  {Mentink}}, \bibinfo {author} {\bibfnamefont {M.~I.}\ \bibnamefont
  {Katsnelson}}, \ and\ \bibinfo {author} {\bibfnamefont {M.}~\bibnamefont
  {Lemeshko}},\ }\href {\doibase 10.1103/PhysRevB.99.064428} {\bibfield
  {journal} {\bibinfo  {journal} {Phys. Rev. B}\ }\textbf {\bibinfo {volume}
  {99}},\ \bibinfo {pages} {064428} (\bibinfo {year} {2019})}\BibitemShut
  {NoStop}%
\bibitem [{\citenamefont {Streib}\ \emph {et~al.}(2019)\citenamefont {Streib},
  \citenamefont {Vidal-Silva}, \citenamefont {Shen},\ and\ \citenamefont
  {Bauer}}]{SVSB19}%
  \BibitemOpen
  \bibfield  {author} {\bibinfo {author} {\bibfnamefont {S.}~\bibnamefont
  {Streib}}, \bibinfo {author} {\bibfnamefont {N.}~\bibnamefont {Vidal-Silva}},
  \bibinfo {author} {\bibfnamefont {K.}~\bibnamefont {Shen}}, \ and\ \bibinfo
  {author} {\bibfnamefont {G.~E.~W.}\ \bibnamefont {Bauer}},\ }\href {\doibase
  10.1103/PhysRevB.99.184442} {\bibfield  {journal} {\bibinfo  {journal} {Phys.
  Rev. B}\ }\textbf {\bibinfo {volume} {99}},\ \bibinfo {pages} {184442}
  (\bibinfo {year} {2019})}\BibitemShut {NoStop}%
\bibitem [{\citenamefont {R\"uckriegel}\ \emph {et~al.}(2020)\citenamefont
  {R\"uckriegel}, \citenamefont {Streib}, \citenamefont {Bauer},\ and\
  \citenamefont {Duine}}]{RSBD20}%
  \BibitemOpen
  \bibfield  {author} {\bibinfo {author} {\bibfnamefont {A.}~\bibnamefont
  {R\"uckriegel}}, \bibinfo {author} {\bibfnamefont {S.}~\bibnamefont
  {Streib}}, \bibinfo {author} {\bibfnamefont {G.~E.~W.}\ \bibnamefont
  {Bauer}}, \ and\ \bibinfo {author} {\bibfnamefont {R.~A.}\ \bibnamefont
  {Duine}},\ }\href {\doibase 10.1103/PhysRevB.101.104402} {\bibfield
  {journal} {\bibinfo  {journal} {Phys. Rev. B}\ }\textbf {\bibinfo {volume}
  {101}},\ \bibinfo {pages} {104402} (\bibinfo {year} {2020})}\BibitemShut
  {NoStop}%
\bibitem [{\citenamefont {Ma}\ \emph {et~al.}(2008)\citenamefont {Ma},
  \citenamefont {Woo},\ and\ \citenamefont {Dudarev}}]{MWD08}%
  \BibitemOpen
  \bibfield  {author} {\bibinfo {author} {\bibfnamefont {P.-W.}\ \bibnamefont
  {Ma}}, \bibinfo {author} {\bibfnamefont {C.~H.}\ \bibnamefont {Woo}}, \ and\
  \bibinfo {author} {\bibfnamefont {S.~L.}\ \bibnamefont {Dudarev}},\ }\href
  {\doibase 10.1103/PhysRevB.78.024434} {\bibfield  {journal} {\bibinfo
  {journal} {Phys. Rev. B}\ }\textbf {\bibinfo {volume} {78}},\ \bibinfo
  {pages} {024434} (\bibinfo {year} {2008})}\BibitemShut {NoStop}%
\bibitem [{\citenamefont {Perera}\ \emph {et~al.}(2016)\citenamefont {Perera},
  \citenamefont {Eisenbach}, \citenamefont {Nicholson}, \citenamefont
  {Stocks},\ and\ \citenamefont {Landau}}]{PEN+16}%
  \BibitemOpen
  \bibfield  {author} {\bibinfo {author} {\bibfnamefont {D.}~\bibnamefont
  {Perera}}, \bibinfo {author} {\bibfnamefont {M.}~\bibnamefont {Eisenbach}},
  \bibinfo {author} {\bibfnamefont {D.~M.}\ \bibnamefont {Nicholson}}, \bibinfo
  {author} {\bibfnamefont {G.~M.}\ \bibnamefont {Stocks}}, \ and\ \bibinfo
  {author} {\bibfnamefont {D.~P.}\ \bibnamefont {Landau}},\ }\href {\doibase
  10.1103/PhysRevB.93.060402} {\bibfield  {journal} {\bibinfo  {journal} {Phys.
  Rev. B}\ }\textbf {\bibinfo {volume} {93}},\ \bibinfo {pages} {060402}
  (\bibinfo {year} {2016})}\BibitemShut {NoStop}%
\bibitem [{\citenamefont {A{\ss}mann}\ and\ \citenamefont
  {Nowak}(2019)}]{AN19}%
  \BibitemOpen
  \bibfield  {author} {\bibinfo {author} {\bibfnamefont {M.}~\bibnamefont
  {A{\ss}mann}}\ and\ \bibinfo {author} {\bibfnamefont {U.}~\bibnamefont
  {Nowak}},\ }\href {\doibase https://doi.org/10.1016/j.jmmm.2018.08.034}
  {\bibfield  {journal} {\bibinfo  {journal} {Journal of Magnetism and Magnetic
  Materials}\ }\textbf {\bibinfo {volume} {469}},\ \bibinfo {pages} {217}
  (\bibinfo {year} {2019})}\BibitemShut {NoStop}%
\bibitem [{\citenamefont {Hellsvik}\ \emph {et~al.}(2019)\citenamefont
  {Hellsvik}, \citenamefont {Thonig}, \citenamefont {Modin}, \citenamefont
  {Iu\ifmmode~\mbox{\c{s}}\else \c{s}\fi{}an}, \citenamefont {Bergman},
  \citenamefont {Eriksson}, \citenamefont {Bergqvist},\ and\ \citenamefont
  {Delin}}]{HTM+19}%
  \BibitemOpen
  \bibfield  {author} {\bibinfo {author} {\bibfnamefont {J.}~\bibnamefont
  {Hellsvik}}, \bibinfo {author} {\bibfnamefont {D.}~\bibnamefont {Thonig}},
  \bibinfo {author} {\bibfnamefont {K.}~\bibnamefont {Modin}}, \bibinfo
  {author} {\bibfnamefont {D.}~\bibnamefont {Iu\ifmmode~\mbox{\c{s}}\else
  \c{s}\fi{}an}}, \bibinfo {author} {\bibfnamefont {A.}~\bibnamefont
  {Bergman}}, \bibinfo {author} {\bibfnamefont {O.}~\bibnamefont {Eriksson}},
  \bibinfo {author} {\bibfnamefont {L.}~\bibnamefont {Bergqvist}}, \ and\
  \bibinfo {author} {\bibfnamefont {A.}~\bibnamefont {Delin}},\ }\href
  {\doibase 10.1103/PhysRevB.99.104302} {\bibfield  {journal} {\bibinfo
  {journal} {Phys. Rev. B}\ }\textbf {\bibinfo {volume} {99}},\ \bibinfo
  {pages} {104302} (\bibinfo {year} {2019})}\BibitemShut {NoStop}%
\bibitem [{\citenamefont {Strungaru}\ \emph {et~al.}(2021)\citenamefont
  {Strungaru}, \citenamefont {Ellis}, \citenamefont {Ruta}, \citenamefont
  {Chubykalo-Fesenko}, \citenamefont {Evans},\ and\ \citenamefont
  {Chantrell}}]{SER+21}%
  \BibitemOpen
  \bibfield  {author} {\bibinfo {author} {\bibfnamefont {M.}~\bibnamefont
  {Strungaru}}, \bibinfo {author} {\bibfnamefont {M.~O.~A.}\ \bibnamefont
  {Ellis}}, \bibinfo {author} {\bibfnamefont {S.}~\bibnamefont {Ruta}},
  \bibinfo {author} {\bibfnamefont {O.}~\bibnamefont {Chubykalo-Fesenko}},
  \bibinfo {author} {\bibfnamefont {R.~F.~L.}\ \bibnamefont {Evans}}, \ and\
  \bibinfo {author} {\bibfnamefont {R.~W.}\ \bibnamefont {Chantrell}},\ }\href
  {\doibase 10.1103/PhysRevB.103.024429} {\bibfield  {journal} {\bibinfo
  {journal} {Phys. Rev. B}\ }\textbf {\bibinfo {volume} {103}},\ \bibinfo
  {pages} {024429} (\bibinfo {year} {2021})}\BibitemShut {NoStop}%
\bibitem [{\citenamefont {Liechtenstein}\ \emph {et~al.}(1987)\citenamefont
  {Liechtenstein}, \citenamefont {Katsnelson}, \citenamefont {Antropov},\ and\
  \citenamefont {Gubanov}}]{LKAG87}%
  \BibitemOpen
  \bibfield  {author} {\bibinfo {author} {\bibfnamefont {A.~I.}\ \bibnamefont
  {Liechtenstein}}, \bibinfo {author} {\bibfnamefont {M.~I.}\ \bibnamefont
  {Katsnelson}}, \bibinfo {author} {\bibfnamefont {V.~P.}\ \bibnamefont
  {Antropov}}, \ and\ \bibinfo {author} {\bibfnamefont {V.~A.}\ \bibnamefont
  {Gubanov}},\ }\href {\doibase 10.1016/0304-8853(87)90721-9} {\bibfield
  {journal} {\bibinfo  {journal} {J. Magn. Magn. Materials}\ }\textbf {\bibinfo
  {volume} {67}},\ \bibinfo {pages} {65} (\bibinfo {year} {1987})}\BibitemShut
  {NoStop}%
\bibitem [{\citenamefont {Polesya}\ \emph {et~al.}(2010)\citenamefont
  {Polesya}, \citenamefont {Mankovsky}, \citenamefont {\v{S}ipr}, \citenamefont
  {Meindl}, \citenamefont {Strunk},\ and\ \citenamefont {Ebert}}]{PMS+10}%
  \BibitemOpen
  \bibfield  {author} {\bibinfo {author} {\bibfnamefont {S.}~\bibnamefont
  {Polesya}}, \bibinfo {author} {\bibfnamefont {S.}~\bibnamefont {Mankovsky}},
  \bibinfo {author} {\bibfnamefont {O.}~\bibnamefont {\v{S}ipr}}, \bibinfo
  {author} {\bibfnamefont {W.}~\bibnamefont {Meindl}}, \bibinfo {author}
  {\bibfnamefont {C.}~\bibnamefont {Strunk}}, \ and\ \bibinfo {author}
  {\bibfnamefont {H.}~\bibnamefont {Ebert}},\ }\href {\doibase
  10.1103/PhysRevB.82.214409} {\bibfield  {journal} {\bibinfo  {journal} {Phys.
  Rev. B}\ }\textbf {\bibinfo {volume} {82}},\ \bibinfo {pages} {214409}
  (\bibinfo {year} {2010})}\BibitemShut {NoStop}%
\bibitem [{\citenamefont {Mryasov}(2005)}]{Mry05}%
  \BibitemOpen
  \bibfield  {author} {\bibinfo {author} {\bibfnamefont {O.~N.}\ \bibnamefont
  {Mryasov}},\ }\href@noop {} {\bibfield  {journal} {\bibinfo  {journal} {Phase
  Transitions}\ }\textbf {\bibinfo {volume} {78}},\ \bibinfo {pages} {197}
  (\bibinfo {year} {2005})}\BibitemShut {NoStop}%
\bibitem [{\citenamefont {Skubic}\ \emph {et~al.}(2008)\citenamefont {Skubic},
  \citenamefont {Hellsvik}, \citenamefont {Nordstr\"om},\ and\ \citenamefont
  {Eriksson}}]{SHNE08}%
  \BibitemOpen
  \bibfield  {author} {\bibinfo {author} {\bibfnamefont {B.}~\bibnamefont
  {Skubic}}, \bibinfo {author} {\bibfnamefont {J.}~\bibnamefont {Hellsvik}},
  \bibinfo {author} {\bibfnamefont {L.}~\bibnamefont {Nordstr\"om}}, \ and\
  \bibinfo {author} {\bibfnamefont {O.}~\bibnamefont {Eriksson}},\ }\href
  {\doibase 10.1088/0953-8984/20/31/315203} {\bibfield  {journal} {\bibinfo
  {journal} {J. Phys.: Cond. Mat.}\ }\textbf {\bibinfo {volume} {20}},\
  \bibinfo {pages} {315203} (\bibinfo {year} {2008})}\BibitemShut {NoStop}%
\bibitem [{\citenamefont {Udvardi}\ \emph {et~al.}(2003)\citenamefont
  {Udvardi}, \citenamefont {Szunyogh}, \citenamefont {Palot\'as},\ and\
  \citenamefont {Weinberger}}]{USPW03}%
  \BibitemOpen
  \bibfield  {author} {\bibinfo {author} {\bibfnamefont {L.}~\bibnamefont
  {Udvardi}}, \bibinfo {author} {\bibfnamefont {L.}~\bibnamefont {Szunyogh}},
  \bibinfo {author} {\bibfnamefont {K.}~\bibnamefont {Palot\'as}}, \ and\
  \bibinfo {author} {\bibfnamefont {P.}~\bibnamefont {Weinberger}},\ }\href
  {\doibase 10.1103/PhysRevB.68.104436} {\bibfield  {journal} {\bibinfo
  {journal} {Phys. Rev. B}\ }\textbf {\bibinfo {volume} {68}},\ \bibinfo
  {pages} {104436} (\bibinfo {year} {2003})}\BibitemShut {NoStop}%
\bibitem [{\citenamefont {Ebert}\ and\ \citenamefont
  {Mankovsky}(2009)}]{EM09a}%
  \BibitemOpen
  \bibfield  {author} {\bibinfo {author} {\bibfnamefont {H.}~\bibnamefont
  {Ebert}}\ and\ \bibinfo {author} {\bibfnamefont {S.}~\bibnamefont
  {Mankovsky}},\ }\href {\doibase 10.1103/PhysRevB.79.045209} {\bibfield
  {journal} {\bibinfo  {journal} {Phys. Rev. B}\ }\textbf {\bibinfo {volume}
  {79}},\ \bibinfo {pages} {045209} (\bibinfo {year} {2009})}\BibitemShut
  {NoStop}%
\bibitem [{\citenamefont {Mankovsky}\ \emph {et~al.}(2020)\citenamefont
  {Mankovsky}, \citenamefont {Polesya},\ and\ \citenamefont {Ebert}}]{MPE20}%
  \BibitemOpen
  \bibfield  {author} {\bibinfo {author} {\bibfnamefont {S.}~\bibnamefont
  {Mankovsky}}, \bibinfo {author} {\bibfnamefont {S.}~\bibnamefont {Polesya}},
  \ and\ \bibinfo {author} {\bibfnamefont {H.}~\bibnamefont {Ebert}},\ }\href
  {\doibase 10.1103/PhysRevB.101.174401} {\bibfield  {journal} {\bibinfo
  {journal} {Phys. Rev. B}\ }\textbf {\bibinfo {volume} {101}},\ \bibinfo
  {pages} {174401} (\bibinfo {year} {2020})}\BibitemShut {NoStop}%
\bibitem [{\citenamefont {Sadhukhan}\ \emph {et~al.}(2022)\citenamefont
  {Sadhukhan}, \citenamefont {Bergman}, \citenamefont {Kvashnin}, \citenamefont
  {Hellsvik},\ and\ \citenamefont {Delin}}]{SBK+22}%
  \BibitemOpen
  \bibfield  {author} {\bibinfo {author} {\bibfnamefont {B.}~\bibnamefont
  {Sadhukhan}}, \bibinfo {author} {\bibfnamefont {A.}~\bibnamefont {Bergman}},
  \bibinfo {author} {\bibfnamefont {Y.~O.}\ \bibnamefont {Kvashnin}}, \bibinfo
  {author} {\bibfnamefont {J.}~\bibnamefont {Hellsvik}}, \ and\ \bibinfo
  {author} {\bibfnamefont {A.}~\bibnamefont {Delin}},\ }\href {\doibase
  10.1103/PhysRevB.105.104418} {\bibfield  {journal} {\bibinfo  {journal}
  {Phys. Rev. B}\ }\textbf {\bibinfo {volume} {105}},\ \bibinfo {pages}
  {104418} (\bibinfo {year} {2022})}\BibitemShut {NoStop}%
\bibitem [{\citenamefont {Ebert}\ \emph {et~al.}(2011)\citenamefont {Ebert},
  \citenamefont {K\"odderitzsch},\ and\ \citenamefont {Min\'{a}r}}]{EKM11}%
  \BibitemOpen
  \bibfield  {author} {\bibinfo {author} {\bibfnamefont {H.}~\bibnamefont
  {Ebert}}, \bibinfo {author} {\bibfnamefont {D.}~\bibnamefont
  {K\"odderitzsch}}, \ and\ \bibinfo {author} {\bibfnamefont {J.}~\bibnamefont
  {Min\'{a}r}},\ }\href {\doibase 10.1088/0034-4885/74/9/096501} {\bibfield
  {journal} {\bibinfo  {journal} {Rep. Prog. Phys.}\ }\textbf {\bibinfo
  {volume} {74}},\ \bibinfo {pages} {096501} (\bibinfo {year}
  {2011})}\BibitemShut {NoStop}%
\bibitem [{Sup()}]{SuppMat}%
  \BibitemOpen
  \href@noop {} {\bibinfo  {journal} {See Supplemental Material for additional
  details on the change of the inverse scattering matrix due to a spin tilting
  and due to a shift of the atom from the equilibrium position, as well as the
  results on three-site spin-lattice interaction parameters calculated in a
  scalar-relativistic way.}\ }\BibitemShut {NoStop}%
\bibitem [{\citenamefont {R\"uckriegel}\ \emph {et~al.}(2014)\citenamefont
  {R\"uckriegel}, \citenamefont {Kopietz}, \citenamefont {Bozhko},
  \citenamefont {Serga},\ and\ \citenamefont {Hillebrands}}]{RKB+14}%
  \BibitemOpen
\bibfield  {journal} {  }\bibfield  {author} {\bibinfo {author} {\bibfnamefont
  {A.}~\bibnamefont {R\"uckriegel}}, \bibinfo {author} {\bibfnamefont
  {P.}~\bibnamefont {Kopietz}}, \bibinfo {author} {\bibfnamefont {D.~A.}\
  \bibnamefont {Bozhko}}, \bibinfo {author} {\bibfnamefont {A.~A.}\
  \bibnamefont {Serga}}, \ and\ \bibinfo {author} {\bibfnamefont
  {B.}~\bibnamefont {Hillebrands}},\ }\href {\doibase
  10.1103/PhysRevB.89.184413} {\bibfield  {journal} {\bibinfo  {journal} {Phys.
  Rev. B}\ }\textbf {\bibinfo {volume} {89}},\ \bibinfo {pages} {184413}
  (\bibinfo {year} {2014})}\BibitemShut {NoStop}%
\bibitem [{\citenamefont {Holstein}\ and\ \citenamefont
  {Primakoff}(1940)}]{HP40}%
  \BibitemOpen
  \bibfield  {author} {\bibinfo {author} {\bibfnamefont {T.}~\bibnamefont
  {Holstein}}\ and\ \bibinfo {author} {\bibfnamefont {H.}~\bibnamefont
  {Primakoff}},\ }\href {\doibase 10.1103/PhysRev.58.1098} {\bibfield
  {journal} {\bibinfo  {journal} {Phys. Rev.}\ }\textbf {\bibinfo {volume}
  {58}},\ \bibinfo {pages} {1098} (\bibinfo {year} {1940})}\BibitemShut
  {NoStop}%
\bibitem [{\citenamefont {B\"ottger}(1983)}]{Boe83}%
  \BibitemOpen
  \bibfield  {author} {\bibinfo {author} {\bibfnamefont {H.}~\bibnamefont
  {B\"ottger}},\ }\enquote {\bibinfo {title} {Principles of the theory of
  lattice dynamics},}\ \ (\bibinfo  {publisher} {Akademie-Verlag},\ \bibinfo
  {address} {Berlin},\ \bibinfo {year} {1983})\BibitemShut {NoStop}%
\end{thebibliography}

%

\end{document}